\documentclass[conference]{IEEEtran}
\IEEEoverridecommandlockouts

\usepackage{cite}
\usepackage{amsmath,amssymb,amsfonts}
\usepackage{algorithmic}
\usepackage{graphicx}
\usepackage{textcomp}
\usepackage{xcolor}
\usepackage{booktabs}
\usepackage{tabularx}
\usepackage{multirow}
\usepackage[hidelinks]{hyperref}
\def\BibTeX{{\rm B\kern-.05em{\sc i\kern-.025em b}\kern-.08em
    T\kern-.1667em\lower.7ex\hbox{E}\kern-.125emX}}

\usepackage{multirow}

\renewcommand{\baselinestretch}{0.945}
\begin{document}

\title{Data Augmentation for\\ Pathological Speech Enhancement\\
\thanks{This work was supported by the Swiss National Science Foundation project 200021\_215187 on ``Pathological Speech Enhancement".}
}

\author{\IEEEauthorblockN{Mingchi Hou$^{1 ,2}$, Enno Hermann$^{1}$, Ina Kodrasi$^{1}$}
\IEEEauthorblockA{
\textit{$^1$Idiap Research Institute, Switzerland} \\
\textit{$^2$École Polytechnique Fédérale de Lausanne (EPFL), Switzerland}\\}

{\tt \{mingchi.hou,enno.hermann,ina.kodrasi\}@idiap.ch}

}

\maketitle
\begin{abstract}
The performance of state-of-the-art speech enhancement (SE) models considerably degrades for pathological speech due to atypical acoustic characteristics and limited data availability. This paper systematically investigates data augmentation (DA) strategies to improve SE performance for pathological speakers, evaluating both predictive and generative SE models. We examine three DA categories, i.e., transformative, generative, and noise augmentation, assessing their impact with objective SE metrics. Experimental results show that noise augmentation consistently delivers the largest and most robust gains, transformative augmentations provide moderate improvements, while generative augmentation yields limited benefits and can harm performance as the amount of synthetic data increases.
Furthermore, we show that the effectiveness of DA varies depending on the SE model, with DA being more beneficial for predictive SE models. While our results demonstrate that DA improves SE performance for pathological speakers, a performance gap between neurotypical and pathological speech persists, highlighting the need for future research on targeted DA strategies for pathological speech.
\end{abstract}


\section{Introduction}
\label{sec: intro}
Speech enhancement (SE) is essential for reliable speech communication in noisy environments, supporting applications such as hearing aids and voice-controlled systems~\cite{oshaughnessy_speech_2024}.
Modern SE approaches such as predictive~\cite{wang_overview_2018} and generative~\cite{richter2023sgmse,jukic2024sb,lu_cdiffuse_2022} approaches typically rely on large deep neural networks trained in a supervised manner, where noisy speech is paired with the corresponding clean speech to guide learning. While these approaches have shown strong performance, their effectiveness typically depends on the availability of substantial training data. Performance can degrade significantly when training data are limited or imbalanced, and models often fail to generalize to unseen noise conditions or speakers~\cite{SE_training_data_size}.

Most SE research to date has focused on neurotypical speakers~\cite{wang_overview_2018, richter2023sgmse, jukic2024sb,lu_cdiffuse_2022}, while speakers with speech disorders have received comparatively little attention~\cite{hou_eusipco_2025,hou2025generalizabilitypredictivegenerativespeech}. Conditions such as hearing impairments, head and neck cancers, or neurodegenerative diseases like Parkinson’s disease (PD) can lead to a wide range of speech disorders~\cite{Darley_dysarthria}. These disorders often affect articulation, phonation, and prosody, and can result in a substantially reduced speech intelligibility~\cite{Darley_dysarthria}.
Importantly, pathological speech exhibits statistical properties that differ from those of neurotypical speech~\cite{PD_ITG,kodrasi_spectro-temporal_2020}, which causes SE models trained exclusively on neurotypical speech to perform poorly on pathological speech~\cite{hou_eusipco_2025,hou2025generalizabilitypredictivegenerativespeech}.
Given the high prevalence of speech disorders~\cite{WHO_2006}, it is critical to develop SE approaches that work effectively for pathological speech. However, training directly on pathological speech datasets is challenging, as publicly available datasets are generally small~\cite{UASpeech2008,Torgo2012,orozco-arroyave_new_2014_pcgita}. Furthermore, several existing pathological datasets are noisy~\cite{icassp2023}, making supervised training based on noisy-clean speech pairs on such datasets infeasible.

To address the challenge of data scarcity and improve the performance of deep learning models, data augmentation (DA) strategies have beenw widely used to expand training datasets. DA has proven effective in various speech processing tasks including
automatic speech recognition~\cite{DA_ASR}, speech separation~\cite{DA_SS}, speech emotion recognition~\cite{DA_SER}, and pathological speech detection~\cite{JAVANMARDI2024_DA_pathology_detection}, where strategies such as noise augmentation, pitch shifting, and time stretching have been incorporated.
However, the application of DA to SE remains largely underexplored. To the best of our knowledge, only~\cite{DA_DNS,Kim2021SpecMixA} investigate the impact of DA strategies on SE performance. While these studies consider augmentation of noise levels and various manipulations in the time-frequency domain, their analysis is limited to a single predictive SE approach on neurotypical speakers.
More recently, zero-shot text-to-speech (TTS) synthesis has been proposed as a way to generate synthetic data for personalized SE systems~\cite{bae2025generativedataaugmentationchallenge}, yet its actual benefit for SE in pathological speech is unknown.

Motivated by this gap in the literature, in this paper we systematically investigate the effectiveness of various DA strategies for pathological SE. We evaluate transformative augmentations (pitch shifting, time stretching, SpecMix~\cite{Kim2021SpecMixA}), generative augmentations (XTTS~\cite{xtts}, YourTTS~\cite{yourtts}), and noise augmentation, using two state-of-the-art SE models, i.e.,  a predictive~\cite{lemercier_storm_2023} and a generative~\cite{jukic2024sb} model. Experiments on the Spanish PC-GITA dataset~\cite{orozco-arroyave_new_2014_pcgita} show that DA effectiveness varies by strategy, augmentation ratio, and SE model. Noise augmentation is overall the most effective strategy, though the optimal augmentation ratio depends on the SE model. Among transformative augmentations, SpecMix yields the most consistent gains, while generative augmentations offer minimal benefit and may degrade performance at high augmentation ratios.

\section{Problem Formulation \& SE Models}
\label{subsec:SE_models}

We consider the noisy microphone signal $y(t)$ at time $t$, i.e., $
y(t) = x(t) + n(t),
$
where $x(t)$ and $n(t)$ denote the clean speech and additive noise signals.
In the short-time Fourier transform (STFT) domain, the noisy microphone signal $Y(k,l)$ at frequency index $k$ and time frame index $l$ is given by
\begin{equation}
Y(k,l) = X(k,l) + N(k,l),
\label{equation:eqSTFT}
\end{equation}
where $X(k,l)$ and $N(k,l)$ are the complex-valued STFT coefficients of the clean speech and noise signals, respectively.
The goal of SE is to estimate the clean speech signal from the noisy observation.
The performance of various SE approaches for pathological speech (i.e., when the clean signal $x(t)$ is produced by pathological speakers) has been systematically compared in~\cite{hou2025generalizabilitypredictivegenerativespeech}.
It has been shown that a predictive complex-valued spectrogram-based regression (CR) approach~\cite{lemercier_storm_2023} and a generative Schr\"{o}dinger Bridge (SB)-based approach~\cite{jukic2024sb} achieve the highest performance, although the performance remains lower than for neurotypical speech. In this paper, we focus on these two approaches and provide a brief overview of them in the following.
For additional details, the interested reader is referred to~\cite{hou2025generalizabilitypredictivegenerativespeech} and the references therein.

\emph{Predictive CR model.} \enspace
In CR, a deep neural network is trained to predict the real and imaginary parts of the clean STFT coefficients from their noisy counterparts \cite{lemercier_storm_2023}. The SE task is formulated as a supervised regression problem, and the model is trained using the mean squared error between the reconstructed time-domain signal $\hat{x}(t)$ (after inverse STFT) and the clean reference signal $x(t)$~\cite{lemercier_storm_2023}.

\emph{Generative SB model.} \enspace
SB is a generative model that constructs an optimal transport path between noisy and clean speech distributions~\cite{jukic2024sb}. Unlike standard forward diffusion which spreads clean data around the noisy observation~\cite{richter2023sgmse}, SB achieves interpolation between clean and noisy speech STFT coefficients using a weighted data prediction loss~\cite{jukic2024sb}.

\section{Augmentation Strategies} \label{subsec:data_aug}

To address the challenge of limited pathological speech data for SE, we investigate the effects of six DA strategies grouped into three categories based on how they modify the data, i.e., transformative, generative, and noise augmentation. Transformative augmentations alter the waveform or its time-frequency representation to increase variability. Generative augmentations synthesize new utterances to expand the dataset. Noise augmentation broadens the range of acoustic conditions without changing the underlying clean speech signals.

\subsection{Transformative Augmentations}

\subsubsection{Pitch shift}
Pitch reflects the harmonic structure and prosodic patterns of speech, both of which can affect the performance of SE. Utterances with low pitch and low pitch variation are often harder to enhance, because their harmonics are tightly spaced and more easily masked by noise~\cite{hou2025influencecleanspeech}. Notably, pitch is one of the commonly affected speech characteristics in pathological speakers~\cite{pitch_change2018}, further motivating the use of pitch shifting as a DA strategy to improve SE performance.
In our implementation, pitch shifting is done using the \emph{librosa} \texttt{pitch\_shift} function\footnote{https://librosa.org/doc/main/generated/librosa.effects.pitch\_shift.html}, which shifts the pitch of a signal by a specified number of semitones $s$, while preserving its temporal structure.

\subsubsection{Time stretch}
Alterations in speech rate are common in many pathological speech conditions, although the specific patterns vary across disorders~\cite{pathology_impact_speech2006}. To generate training samples covering a wider range of speech rates, we employ time stretching as a DA strategy.
In our implementation, time streching is done using the \emph{librosa} \texttt{time\_strech} function\footnote{https://librosa.org/doc/main/generated/librosa.effects.time\_stretch.html}, which modifies the speech rate by scaling the duration of the audio signal with a ratio $r$, while preserving its pitch.

\subsubsection{SpecMix}

SpecMix aims to improve the generalizability of SE models by generating new training spectrograms through mixing two existing spectrograms using time-frequency masks~\cite{Kim2021SpecMixA}.
In our implementation\footnote{https://github.com/GT-KIM/specmix},  we mix up to three frequency bands and up to three time bands, with the width of each band controlled by the parameter $\gamma$.

\subsection{Generative Augmentations}

To synthesize new utterances, we select two pretrained TTS models that support Spanish (since our SE experiments in Section~\ref{sec: exp} use a Spanish pathological speech dataset)
and have zero-shot voice cloning capabilities.

\subsubsection{YourTTS}

YourTTS \cite{yourtts} is an extension of the Vits architecture~\cite{vits} and
trained end-to-end with a HiFi-GAN~\cite{hifigan} vocoder. Language embeddings
allow multilingual synthesis and an external speaker encoder model creates
speaker embeddings from reference audio, including for unseen speakers.

\subsubsection{XTTS}

XTTS \cite{xtts} tokenizes text via byte-pair encoding and speech via vector
quantization to enable large-scale training. An integrated speaker encoder
computes a speaker embedding from reference audio. Then, a GPT-2 model predicts output
speech tokens before a HiFi-GAN vocoder generates the final waveform.

\subsection{Noise Augmentation}
Noise augmentation is a common DA strategy which is typically used to increase the quantity and diversity of training data by corrupting clean signals with noise~\cite{DA_ASR, DA_SER, DA_SS, JAVANMARDI2024_DA_pathology_detection}. In SE, the training set already consists of noisy-clean speech. We use the term noise augmentation to refer to expanding the existing training set by generating additional noisy–clean pairs through considering more input signal-to-noise ratios (SNRs), without altering the original clean or noise signals.

\section{Experimental Settings}\label{sec:experimental_settings}

\subsection{Dataset}\label{subsec:datasets}

{\emph{Clean speech.}} \enspace
We use the Spanish PC-GITA dataset~\cite{orozco-arroyave_new_2014_pcgita}, which contains $2.8$ hours of recordings from $50$ neurotypical and $50$ pathological speakers diagnosed with PD. Each speaker utters $12$ utterances ($10$ sentences, one text passage, a spontaneous monologue). Prior to further processing, recordings are downsampled from $44.1$ kHz to $16$ kHz.

\emph{Noisy mixtures.} \enspace
To generate noisy mixtures, we use noise recordings sampled at $16$ kHz from the CHiME3 dataset~\cite{barker_chime3_2015}. For each clean utterance, we randomly select from four noise types \{BUS, CAF, STR, PED\}, with an SNR chosen uniformly from the interval [$-6$, $14$] dB.

\subsection{Training}\label{subsec:training_configurations}

Due to the limited size of the PC-GITA dataset, which is a common characteristic of pathological speech datasets\footnote{For comparison, large-scale SE datasets such as WSJ0~\cite{garofolo_john_s_csr-i_2007_wsj0} contain $28.6$ hours of speech.}, we conduct a $10$-fold speaker-independent cross-validation. Within each fold, data are partitioned into $80$\% for training, $10$\% for validation, and $10$\% for testing.

As in~\cite{lemercier_storm_2023}, the signals are represented in the STFT domain with a window of $510$ samples and a hop size of $128$ samples. The resulting spectrogram undergoes dynamic range compression controlled by the parameters $\alpha = 0.5$ and $\beta = 0.33$, consistent with the approach in~\cite{jukic2024sb}.
Both CR and SB adopt the NCSN++ backbone implemented in U-Net style architecture with ResNet blocks~\cite{song2019generative}, further modified for complex inputs as in~\cite{lemercier_storm_2023}.
The SB model uses
extra noise scheduling layers, incorporating a variance-exploding schedule with hyperparameters $\sigma_{\min}=0.7$, $\sigma_{\max}=1.82$, and $50$-step SDE sampling. The number of trainable parameters for CR and SB is $22.1$ million and $25.2$ million, respectively.

Training is carried out using the Adam optimizer with a batch size of $4$, a learning rate of $10^{-4}$, and a maximum of $1000$ epochs. Training stops if the validation loss does not decrease for $20$ consecutive epochs. Training and inference for the CR and SB model are performed on an RTX 3090 GPU and an NVIDIA H100 GPU, respectively.

\subsection{Augmentation}
To enable a systematic comparison between each DA strategy, we use the following settings.

{\emph{Transformative augmentations. \enspace}}
As in~\cite{JAVANMARDI2024_DA_pathology_detection}, we generate four pitch shifted versions of each clean utterance in the dataset using $s \in \{-2.5, -1.5, 1.5, 2.5\}$ semitones. Similarly, four time stretched versions are created using $r \in \{0.81, 0.93, 1.07, 1.23\}$. Finally, four SpecMix variants are generated following~\cite{Kim2021SpecMixA}, with a fixed $\gamma \in \{0.1, 0.3, 0.5\}$ and an additional $\gamma$ value sampled uniformly from $[0,1]$.
For each generated utterance, a noise type is randomly selected and mixed at a randomly chosen SNR.
We evaluate each DA strategy at three augmentation ratios, i.e., $25\%$, $100\%$, and $400\%$, indicating the fraction of augmented data added relative to the original dataset. For the $400\%$ ratio, all generated augmented utterances are used. For the $25\%$ and $100\%$ ratios, subsets of the generated augmented data are randomly sampled for each speaker.

{\emph{Generative augmentations.}} \enspace
For generative augmentation, we synthesize entirely new clean utterances using YourTTS and XTTS.
For YourTTS, we use a model trained on $3200$~hours of the CML-TTS
dataset~\cite{cml-tts}, including $279$~hours of Spanish speech. For XTTS, we use
the v2 model, which is trained on over $27000$~hours of speech from Common
Voice~\cite{commonvoice} and private datasets, including $1500$~hours of Spanish speech.
We run both models with the Coqui TTS
toolkit\footnote{\url{https://github.com/idiap/coqui-ai-TTS}} and default
inference settings. 
For each speaker, one reference sentence from the PC-GITA training set is used for voice cloning. Using this reference, $12$ synthetic utterances are generated per speaker for each TTS model.
This corresponds to using an augmentation ratio of 100\%, since the original PC-GITA dataset contains $12$ utterances per speaker. 
To evaluate the 25\% augmentation ratio, subsets of the synthetic utterances are randomly selected per speaker. 
It should be noted that we do not consider an augmentation ratio of 400\%, as preliminary experiments showed that generating more synthetic data did not yield further improvements, while substantially increasing computational cost.
As with transformative augmentations, noisy versions of each synthetic utterance are generated by mixing with a randomly selected noise type at a random SNR.

{\emph{Noise augmentation.}} \enspace For noise augmentation, additional noisy versions are generated from the original clean signals by combining each clean utterance with a randomly selected noise type at a randomly chosen SNR (i.e., following the same procedure as in the original training set). For the 400\% augmentation ratio, four noisy versions are generated per clean file. For the 25\% and 100\% ratios, subsets of these noisy versions are randomly selected per speaker to match the target augmentation ratio.

\subsection{Evaluation}\label{subsec:evaluation_metrics}

The performance of SE is evaluated using the wideband perceptual evaluation of speech quality (PESQ)~\cite{pesq_2001} and the frequency-weighted segmental SNR (fwSSNR)~\cite{fwSSNR} metrics. The clean speech signal is used as the reference signal.
Performance improvement in these metrics, i.e., $\Delta$PESQ and $\Delta$fwSSNR, is computed as the difference between the PESQ and fwSSNR values of the enhanced signal and the corresponding noisy microphone signal.

\section{Results and Discussion}
\label{sec: exp}

In this section, we present the results of our evaluation of different DA strategies for SE. We first examine each augmentation category separately for the CR and SB models in Sections~\ref{sec: aug_cr} and~\ref{sec: aug_sb}, highlighting how individual strategies and augmentation ratios affect performance for pathological speakers. We then provide an overall comparison that includes neurotypical speakers in Section~\ref{sec: aug_all}, allowing us to quantify the remaining performance gap of SE for pathological speech.

\subsection{DA Strategies for the CR Model on Pathological Speakers} 
\label{sec: aug_cr}

\begin{figure}[t!]
\centering
\includegraphics[width=0.47\textwidth]{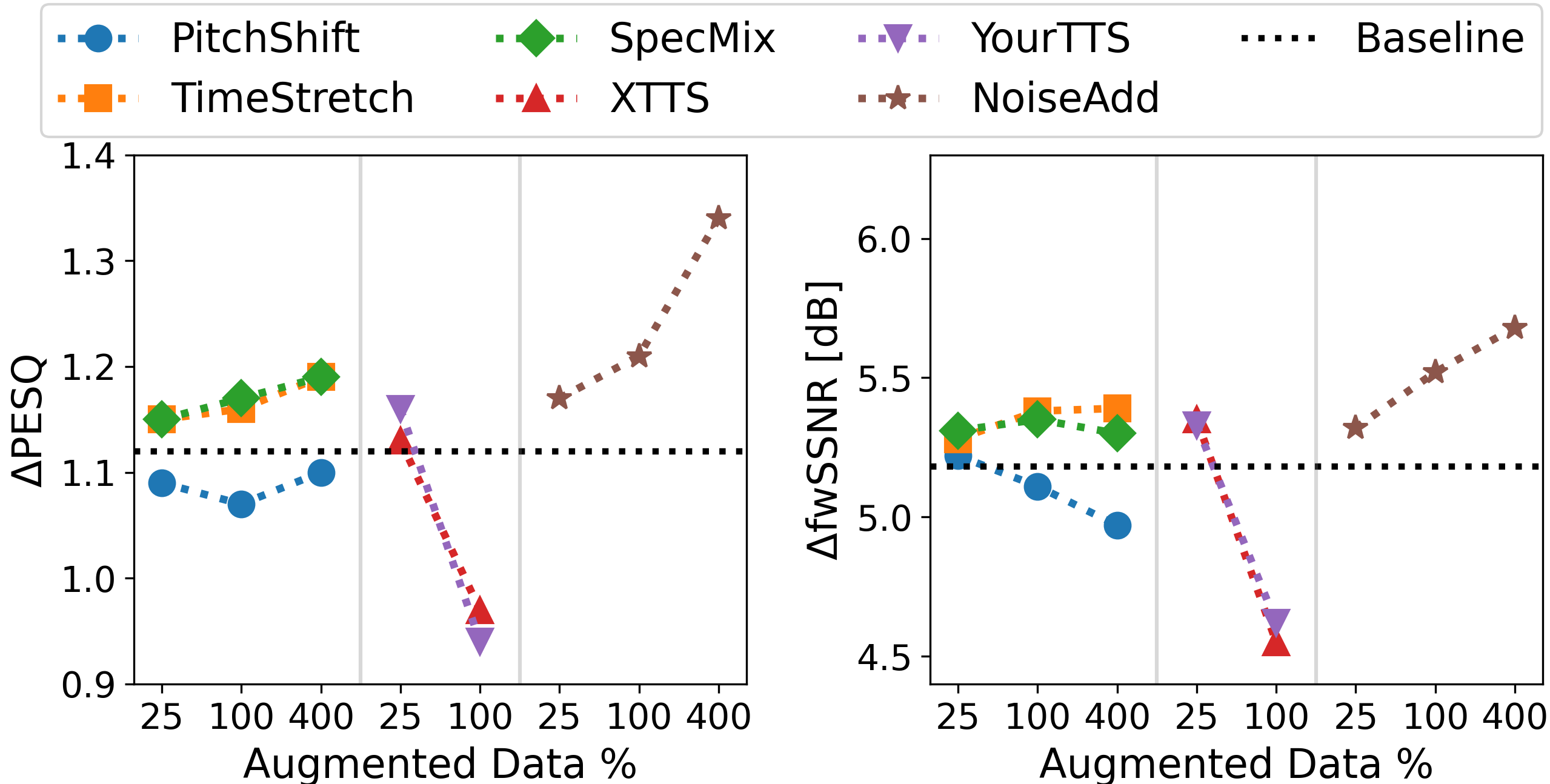}
  \caption{$\Delta$PESQ (left) and $\Delta$fwSSNR (right) for pathological speakers using the CR model with strategies from three DA categories at different augmentation ratios.  Within each category, separate plots correspond to individual strategies. For reference, the baseline performance without any DA strategy is also shown.
  }
  \label{fig:CR}
\end{figure}

Fig.~\ref{fig:CR} shows the $\Delta$PESQ and $\Delta$fwSSNR for the CR model on pathological speakers across all DA strategies and augmentation ratios. For reference, the baseline performance without DA is also presented.

For transformative augmentations, both metrics indicate that modest performance improvements can be achieved with time stretching and SpecMix, whereas pitch shifting degrades performance compared to the baseline. 
Further, it can be observed that moderate augmentation ratios generally provide consistent improvements with time stretching and SpecMix, while aggressive augmentation can lead to performance saturation (e.g., in $\Delta$fwSSNR at a ratio of 400\%).

For generative augmentations, both metrics show that regardless of the TTS model used, only minimal performance improvements are obtained at a small augmentation ratio of 25\%. Increasing the augmentation ratio to 100\% leads to a substantial performance degradation compared to the baseline. We argue that this occurs because the TTS models were trained on neurotypical speech and therefore do not closely resemble the reference pathological speech characteristics. This observation is supported by our informal listening tests.

For noise augmentation, both metrics show considerable improvements over the baseline for all augmentation ratios, with higher augmentation levels producing progressively greater gains. We attribute this to the model learning to be more robust to additive distortions, allowing it to better separate the clean speech components from noise.

Overall, noise augmentation consistently provides the largest gains across all ratios, considerably outperforming all other strategies. Moderate time stretching and SpecMix offer modest improvements, while pitch shifting and generative augmentations degrade performance.

\subsection{DA Strategies for the SB model on Pathological Speakers}
\label{sec: aug_sb}
\begin{figure}[t!]
\centering
\includegraphics[width=0.47\textwidth]{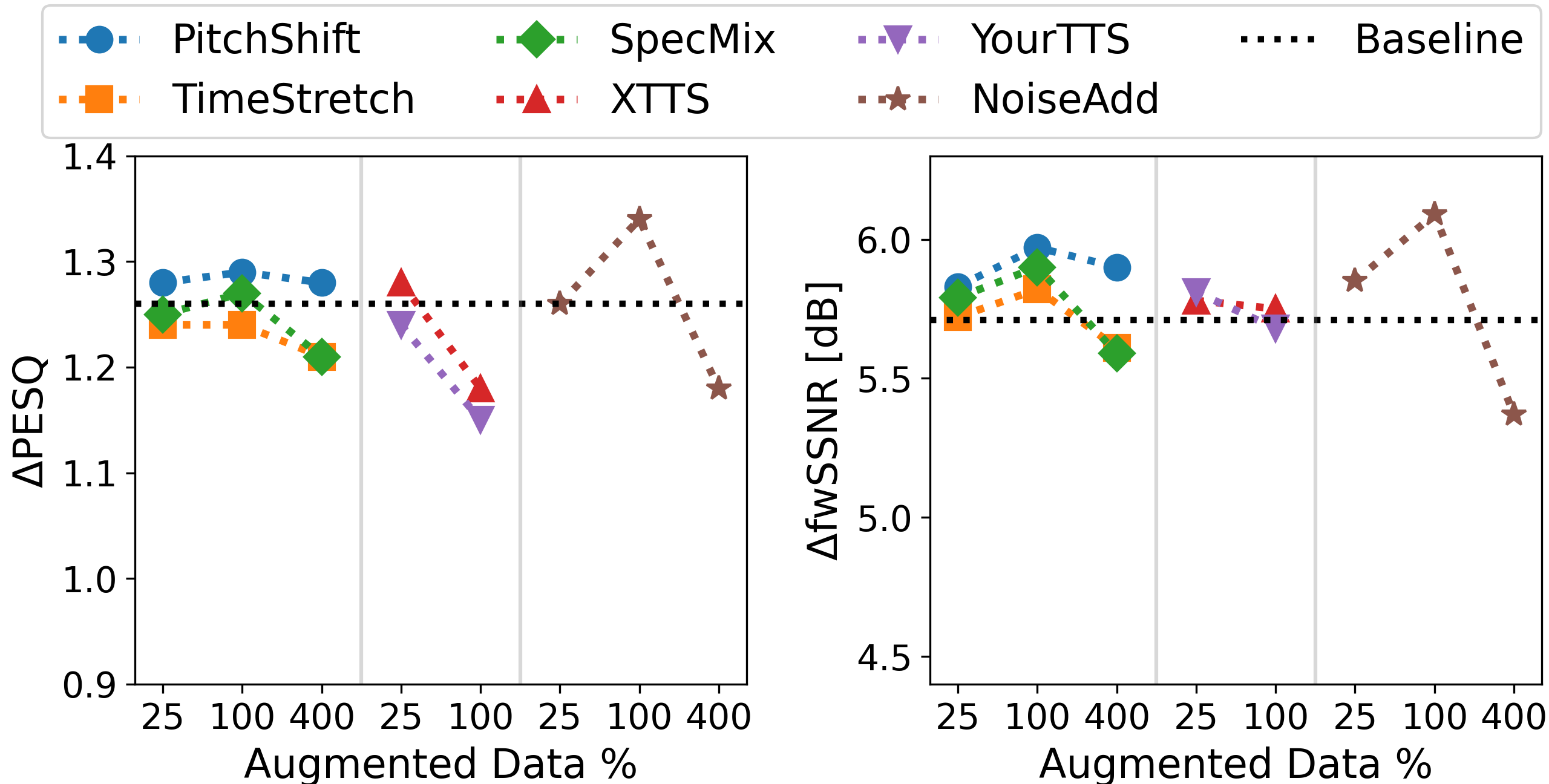}
  \caption{$\Delta$PESQ (left) and $\Delta$fwSSNR (right) for pathological speakers using the SB model with strategies from three DA categories at different augmentation ratios.  Within each category, separate plots correspond to individual strategies. For reference, the baseline performance without any DA strategy is also shown.
  }
  \label{fig:SB}
\end{figure}

Fig.~\ref{fig:SB} shows the $\Delta$PESQ and $\Delta$fwSSNR for the SB model on pathological speakers across all DA strategies and augmentation ratios, with the baseline performance without DA included for reference.

For transformative augmentations, both metrics indicate that modest improvements can be obtained with pitch shifting and SpecMix, whereas time stretching may deteriorate performance. This contrasts with the CR model, where time stretching was beneficial and pitch shifting was harmful, highlighting that the effectiveness of transformative augmentations critically depends on the training objective. Similarly to the CR model, moderate augmentation ratios provide minor improvements, while aggressive ratios can degrade performance. 

As with the CR model, both metrics indicate that generative augmentations offer no benefit for the SB model and can cause considerable performance degradation at 100\% augmentation.

\begin{table*}[t!]
    \centering
    \caption{$\Delta$fwSSNR (mean $\pm$ confidence interval) for neurotypical and pathological speakers using the CR and SB models. Baseline performance (no DA) and performance for the most effective strategy from each DA category is presented. 
    Values in bold indicate the highest performance improvement obtained for each group and each model.}
    \label{tbl: comparison}
    \addtolength{\tabcolsep}{-0.4em}
\begin{tabularx}{\textwidth}{X|cccc|cccc}
    \toprule
      & \multicolumn{4}{c|}{CR model} & \multicolumn{4}{c}{SB model} \\
    Speaker  & Baseline & SpecMix ($100$\%) & XTTS ($25$\%) &  NoiseAdd ($400$\%) & Baseline & SpecMix ($100$\%) & XTTS ($25$\%) & NoiseAdd ($100$\%)\\
    \midrule
     Neurotypical & $  5.73 \pm 0.15 $ &  $  5.93 \pm 0.14 $ & $  5.92 \pm 0.14 $ & $ \bf 6.37 \pm 0.14 $ & $  6.47 \pm 0.14$ & $  6.67 \pm 0.15$ & $ 6.51 \pm 0.14$ & $ \bf 6.91 \pm 0.14$ \\
     Pathological& $5.18 \pm 0.18$ & $5.35 \pm 0.16$ & $5.35 \pm 0.16$ &  $ \bf 5.68 \pm 0.17$ & $ 5.71 \pm 0.17$ & $5.90 \pm 0.17$ & $5.78 \pm 0.17$ &  $ \bf 6.09 \pm 0.17$ \\
    \bottomrule
    \end{tabularx}
\end{table*}

For noise augmentation, moderate augmentation ratios improve performance by exposing the SB model to diverse noisy conditions during training. However, differently from the CR model, an aggressive augmentation ratio of 400\% can degrade performance. We argue that this occurs because the SB model relies on learning smooth conditional distributions of clean speech trajectories. Too many noisy variants for the same clean target increases the variability in the conditional input distribution, making it harder for the model to learn consistent generative paths.

These results indicate that noise augmentation with a moderate ratio of $100\%$ is the most effective approach for generative SB-based SE for pathological speech.

\subsection{DA Strategies on Neurotypical and Pathological Speakers}
\label{sec: aug_all}
As noted in Section~\ref{sec: intro}, prior work on DA for SE has been limited to a small set of transformative strategies and a single predictive model. Building on the pathological speech results presented in Sections~\ref{sec: aug_cr} and~\ref{sec: aug_sb}, this section compares the performance of DA strategies for both neurotypical and pathological speakers, providing a foundation for evaluating the remaining performance gap in pathological speech SE. For this analysis, we select the highest-performing strategy from each augmentation category, i.e., SpecMix at an augmentation ratio of $100\%$, XTTS at $25\%$, and noise augmentation (at $400\%$ for CR and $100\%$ for SB).

Table~\ref{tbl: comparison} presents $\Delta$fwSSNR for both SE models across the selected augmentation strategies, reported separately for neurotypical and pathological speakers. Overall, the relative performance trends are consistent across both speaker groups, i.e., SpecMix and XTTS provide only minor improvements for both models, whereas the largest improvements are achieved through noise augmentation. Although the effectiveness of DA strategies is similar for both groups of speakers, a performance gap remains between neurotypical and pathological speakers even for the best performing DA strategy of noise augmentation. These results underscore the need for further research to develop DA strategies that more effectively address the unique characteristics of pathological speech.

\section{Conclusion}
In this paper we systematically investigated the effectiveness of three categories of DA strategies (i.e., transformative, generative, and noise augmentation) to improve the performance of predictive and generative SE approaches for pathological speakers. Our experiments demonstrated that DA can improve SE performance for pathological speakers, but its effectiveness depends on the type of augmentation, the augmentation ratio, and the SE model. While noise augmentation consistently yields robust gains and transformative strategies offer moderate improvements,  generative augmentations can degrade performance if overused, highlighting that more synthetic data does not always lead to better results.
Despite performance improvements, a performance gap with neurotypical speech remains, underscoring the need for DA strategies particularly tailored to pathological speech.

\bibliographystyle{IEEEtran}
\bibliography{mybib}

\end{document}